\documentclass[a4paper,10pt]{article}

\usepackage[colorlinks=true, linktoc=page, linkcolor=blue, citecolor=blue, urlcolor=blue]{hyperref}

\usepackage[scale=0.76,vmarginratio={3:4}]{geometry}

\usepackage[utf8]{inputenc}
\usepackage[T1]{fontenc}

\usepackage{lmodern}
\usepackage{textcomp}
\usepackage{microtype}

\usepackage{mathtools}
\usepackage{amssymb}
\usepackage{mathrsfs}
\usepackage{physics}
\usepackage{units}
\usepackage{bm}
\usepackage{isotope}

\makeatletter
\g@addto@macro\bfseries{\boldmath}
\makeatother

\usepackage{enumitem}
\usepackage[font=small]{caption}
\usepackage{booktabs}
\usepackage{multirow}

\usepackage{authblk}

\usepackage[sorting=none,style=numeric-comp, giveninits=true]{biblatex}

\DeclareSourcemap{\maps[datatype=bibtex,overwrite=true]{
\map{\step[fieldsource=collaboration,final]\step[fieldset=usera,origfieldval,final]}
\map{\step[fieldsource=reportNumber,final]\step[fieldset=userb,origfieldval,final]}
}}

\renewbibmacro*{author}{\iffieldundef{usera}{\printnames{author}}{\textbf{\printfield{usera}}}}
\renewbibmacro*{in:}{\iffieldundef{journal}{}{\printtext{\bibstring{in}\intitlepunct}}}
\renewbibmacro*{finentry}{\iffieldundef{userb}{}{\textls[0]{\small{\textsc{\newunitpunct\textnumero\intitlepunct\printfield{userb}}}}}\finentry}

\bibliography{New_Physics_and_Heavy_ions-EPPS}
\newcommand{\gaga}{{\gamma\gamma}}
\newcommand{\epem}{e^+e^-}

\newcommand{\sqrtsnn}{\sqrt{s_{_{\mbox{\rm \tiny{NN}}}}}}
\newcommand{\ie}{i.e.}
\newcommand{\eg}{e.g.}
\newcommand{\Lumi}{\mathcal L}

\title{New physics searches with heavy-ion collisions at the CERN Large Hadron Collider}

\author[1]{Roderik~Bruce}
\affil[1]{CERN, BE Department, CH-1211 Geneva, Switzerland}

\author[2]{David~d'Enterria\footnote{Corresponding author: david.d'enterria@cern.ch}}
\affil[2]{CERN, EP Department, CH-1211 Geneva, Switzerland}

\author[2]{Albert~de Roeck}

\author[3]{Marco~Drewes}
\affil[3]{Centre for Cosmology, Particle Physics and Phenomenology,\protect\\Universit\'e catholique de Louvain, B-1348 Louvain-la-Neuve, Belgium}

\author[4]{Glennys~R.~Farrar}
\affil[4]{Centre for Cosmology and Particle Physics, New York University, NY, NY 10003, USA}

\author[3]{Andrea~Giammanco}

\author[5]{Oliver~Gould}
\affil[5]{Helsinki Institute of Physics, University of Helsinki, FI-00014, Finland}

\author[3]{Jan~Hajer}

\author[6]{Lucian~Harland-Lang}
\affil[6]{Rudolf Peierls Centre, Beecroft Building, Parks Road, Oxford, OX1 3PU, UK}

\author[3]{Jan~Heisig}
\author[1]{John~M.~Jowett}

\author[7]{Sonia~Kabana\footnote{Current address: Instituto de Alta Investigaci\'on, Universidad de Tarapac\'a, Casilla 7D, Arica, Chile.}}

\affil[7]{Department of Physics, University of Nantes, and SUBATECH, 4 rue Alfred Kastler, 44307 Nantes, France}

\author[3]{Georgios~K.~Krintiras\footnote{Current address: Univ. of Kansas, Dept. Physics \& Astronomy, 1251 Wescoe Hall Dr., Lawrence, KS 66045-7582, USA
}}

\author[8,9,10]{Michael~Korsmeier}
\affil[8]{Institute for Theoretical Particle Physics and Cosmology, RWTH Aachen University, Aachen, Germany}
\affil[9]{Dipartimento di Fisica, Universit\`a di Torino, Torino, Italy}
\affil[10]{Istituto Nazionale di Fisica Nucleare, Sezione di Torino, Torino, Italy}

\author[3]{Michele~Lucente}

\author[11,12]{Guilherme~Milhano}
\affil[11]{LIP, Av. Prof. Gama Pinto, 2, P-1649-003 Lisboa, Portugal}
\affil[12]{Instituto Superior T\'ecnico, Universidade de Lisboa, Av. Rovisco Pais 1, 1049-001, Lisboa, Portugal}

\author[13]{Swagata~Mukherjee}
\affil[13]{III.\ Physikalisches Institut A, RWTH Aachen University, Aachen, Germany}

\author[2]{Jeremi~Niedziela}

\author[14]{Vitalii~A.~Okorokov}
\affil[14]{National Research Nuclear University MEPhI, Moscow, Russia}

\author[15]{Arttu~Rajantie}
\affil[15]{Department of Physics, Imperial College London, London SW7 2AZ, UK}

\author[1]{Michaela~Schaumann}

\date{December 18, 2018}

\begin{document}

\maketitle

\begin{abstract}
This document summarises proposed searches for new physics accessible in the heavy-ion mode at the CERN Large Hadron Collider (LHC), both through hadronic and ultraperipheral $\gaga$ interactions, and that have a competitive or, even, unique discovery potential compared to standard proton-proton collision studies. Illustrative examples include searches for new particles --- such as axion-like pseudoscalars, radions, magnetic monopoles, new long-lived particles, dark photons, and sexaquarks as dark matter candidates --- as well as new interactions, such as non-linear or non-commutative QED extensions. 
We argue that such interesting possibilities constitute a well-justified scientific motivation, complementing standard quark-gluon-plasma physics studies, to continue running with ions at the LHC after the Run-4, i.e., beyond 2030, including light and intermediate-mass ion species, accumulating nucleon-nucleon integrated luminosities in the accessible $\unit{fb^{-1}}$ range per month.
\end{abstract}

\vspace{4.cm}

\thispagestyle{empty}

\pagebreak

\tableofcontents

\setcounter{page}{1}

\section{Introduction}

Physics beyond the Standard Model (SM) is necessary in order to explain numerous unsolved empirical and theoretical problems in high energy physics (see \eg~\cite{CidVidal:2018eel} for a recent review).
Prominent examples among them are the nature of dark matter (DM), the origin of matter-antimatter asymmetry (baryogenesis), and finite neutrino masses, on the one hand, as well as the Higgs mass fine-tuning, the null $\theta_\text{QCD}$ Charge-Parity (CP) violation term in Quantum Chromodynamics (QCD), the origin of fermion families and mixings, charge quantisation, the cosmological constant, and a consistent description of quantum gravity, on the other hand.
Most solutions to these problems require new particles --- such as supersymmetric partners, dark photons, right-handed neutrinos, axions, monopoles --- 
and/or new interactions, which have so far evaded observation due to their large masses and/or their small couplings to SM particles.
Two common complementary routes are followed at colliders in order to search for beyond Standard Model (BSM) physics.
If BSM appears at high masses, one needs to maximise the center-of-mass (c.m.) energy $\sqrt s$.
If BSM involves small couplings, one needs to maximise the luminosity $\Lumi$.
At face value, both strategies present obvious drawbacks for searches in heavy-ion (HI) compared to $pp$ collisions at the 
Large Hadron Collider (LHC):
(i) PbPb collisions run at roughly 2.5 times lower nucleon-nucleon c.m.\ energies than $pp$ collisions ($\sqrtsnn = \unit[5.5\text{ vs.\ }14]{TeV}$), and
%$2.5 \cdot \sqrt s$ (\unit[5.5]{TeV} compared to \unit[14]{TeV}), and
(ii) the nucleon-nucleon luminosities are about a factor 100 smaller (in 2018, $\Lumi_\text{NN} = A^2 \times 6 \cdot \unit[10^{27}]{cm^{-2} s^{-1}} = \unit[2.5 \cdot 10^{32}]{cm^{-2} s^{-1}}$ for PbPb vs.\ $\Lumi_\text{pp} = \unit[2 \cdot 10^{34}] {cm^{-2} s^{-1}}$).

During the high-luminosity Large Hadron Collider (HL-LHC) phase~\cite{Apollinari:2015bam,Apollinari:2017cqg}, whose main focus is BSM searches, the luminosity of $pp$ collisions will be maximised, inevitably leading to a large number of overlapping collisions per bunch crossing (pileup). Pileup translates into a rising difficulty to record all interesting $pp$ events, and thereby an unavoidable increase of the kinematical thresholds for triggers and reconstruction objects in order to reduce unwanted backgrounds. 
Pileup also leads to an intrinsic complication in the reconstruction of exclusive final-states (in particular neutral ones, such as $X\to\gaga$ decays at not too high masses) and of displaced vertices from \eg\ long-lived-particles (LLPs) that appear in many BSM scenarios.
In this context, if BSM has low couplings with the SM and is ``hiding well'' at relatively low masses with moderately ``soft'' final states, HI collisions --- with negligible pileup, optimal primary vertexing (thanks to the large number of primary tracks), reduced trigger thresholds (down to zero $p_T$, in some cases), plus unique and ``clean'' $\gamma\gamma$ exclusive final-states in ultraperipheral interactions~\cite{Baltz:2007kq} with luminosities enhanced by factors of order $\Lumi_\text{PbPb}(\gamma\gamma)/\Lumi_{pp}(\gamma\gamma) = Z^4 \times\Lumi_\text{PbPb}/\Lumi_{pp} = 4.5 \cdot 10^{7} \times (6 \cdot 10^{27})/(2 \cdot 10^{34}) \approx 10$ --- present clear advantages compared to $pp$.

The purpose of this document is to summarise various novel BSM search possibilities accessible at the LHC in the HI mode, and thereby provide new arguments  that strengthen the motivations to prolong the HI programme beyond the LHC Run-4 (\ie\ after 2029). 
A selection of new physics (NP) searches that are competitive with (or, at least, complementary to)  $pp$ studies at the LHC are listed in Table~\ref{tab:BSM_particles}, and succinctly presented hereafter. This list is not comprehensive, but is representative of the type of processes that are attractive and accessible with ions from the perspective of BSM searches. After a summary of the LHC heavy-ion performance of current and future runs (Section~\ref{sec:accelerator}), the document is organised along the following four BSM production mechanisms: 
\begin{enumerate}[label=\arabic*)]
\item Ultraperipheral $\gamma\gamma$ collisions (UPCs), producing, \eg, axion-like particles (ALPs), Section~\ref{sec:UPC}.
\item ``Schwinger'' production through strong classical EM fields, producing, \eg, monopoles, Section~\ref{sec:EM}.
\item Hard scattering processes, producing, \eg, displaced signals from new LLPs, Section~\ref{sec:hard_scatt}.
\item Thermal production in the quark-gluon-plasma (QGP), producing, \eg, sexaquarks, Section~\ref{sec:thermal}.
\end{enumerate}
Processes 1), 2), and 4) explicitly use a BSM production mechanism that is unique in HI collisions (or significantly enhanced compared to the $pp$ mode), whereas in processes of the type 3), it is the comparatively reduced pileup backgrounds that renders HI collisions interesting. 
%\ifnotarxiv
In addition, detailed studies in proton-nucleus and light-ion collisions are needed as a baseline for astrophysics BSM searches, as well as to explain several anomalies observed in ultra-high-energy cosmic ray data (Section~\ref{sec:CR}).
%\fi

\begin{table}
\centering
\begin{tabular}{lll} \toprule
Production mode &
BSM particle/interaction &
Remarks\\ \midrule
\multirow{4}{*}{Ultraperipheral} &
Axion-like particles &
$\gamma\gamma\to a$, $m_a \approx \unit[0.5\text{--}100]{GeV}$ \\
% ultraperipheral
&
Radion &
$\gamma\gamma\to \phi$, $m_\phi \approx \unit[0.5\text{--}100]{GeV}$ \\
% ultraperipheral &
%graviton &
% $\gamma\gamma\to G $, $m_G\approx 0.5$--100~GeV\\
% ultraperipheral &
%monopolonium &
% $\gamma\gamma\to [M\bar{M}] $, \\
% ultraperipheral &
%Other BSM onia &
% $\gamma\gamma\to [M\bar{M}]$ monopolonium, $\gamma\gamma\to S[\tilde{t}\bar{\tilde{t}}]$ stoponium \\
% ultraperipheral &
%unparticles &
% $\gamma\gamma\to\gamma\gamma$ \\
% ultraperipheral
&
Born-Infeld QED &
via $\gamma\gamma\to\gamma\gamma$ anomalies \\
% ultraperipheral
&
Non-commutative interactions &
via $\gamma\gamma\to\gamma\gamma$ anomalies \\ \midrule
% ultraperipheral &
%SUSY &
% $\gamma\gamma\to \tilde{f}\bar{\tilde{f}},\ \tilde{\chi}^+_i\tilde{\chi}^-_i$, Low-mass (excluded by LEP?)\\\midrule
%SUSY sfermions &
% Low-mass (excluded by LEP?)\\
% $\gamma\gamma\to \tilde{f}\bar{\tilde{f}},\ \tilde{\chi}^+_i\tilde{\chi}^-_i$ &
%SUSY gauginos &
% $\gamma\gamma\to \tilde{g}\tilde{g} $ &
% Low-mass (excluded by LEP?)\\
%SUSY (pseudos)scalar Higgs &
% $\gamma\gamma \to H,A\to b\bar b$ &
% Low-mass (excluded by LEP?)\\
%SUSY charged Higgs &
% $\gamma\gamma\to H^+H^-$ &
% Low-mass (excluded by LEP?)\\
Schwinger process &
Magnetic monopole &
Only viable in HI collisions\\ \midrule
\multirow{2}{*}{Hard scattering} &
Dark photon &
$m_{A'} \lesssim \unit[1]{GeV}$, advanced particle ID\\
% hard scattering
&
Long-lived particles (heavy $\nu$) &
$m_\text{LLP} \lesssim \unit[10]{GeV}$, improved vertexing\\\midrule
%\multirow{2}{*}{Thermal QCD} &
Thermal QCD &
Sexaquarks &
DM candidate\\
% thermal QCD
%&
%Strangelets &\\ 
\bottomrule
\end{tabular}
\caption{Examples of new-physics particles and interactions accessible in searches with HI collisions at the LHC, listed by production mechanism. Indicative competitive mass ranges and/or the associated measurement advantages compared to the $pp$ running mode are given.}
\label{tab:BSM_particles}
\end{table}

\section{Accelerator considerations}
\label{sec:accelerator}

The nominal LHC operation schedule includes HI collisions during typically  one month each year, and even when accounting for the roughly  $\times$10 lower integrated running time than $pp$, several BSM searches appear more competitive with ions than with protons as shown below. 
The performance of the HI runs up until the end of Run-2 has been very good, reaching instantaneous PbPb luminosities six times higher than the design value of $\unit[10^{27}]{cm^{-2} s^{-1}}$ (equivalent to a nucleon-nucleon luminosity of $\Lumi_\mathrm{NN}=\unit[2.5 \cdot 10^{32}]{cm^{-2} s^{-1}}$). Four LHC experiments are now taking data with HI collisions, and physics runs have also been carried out with a novel mode of operation with $p$Pb collisions that was not initially foreseen~\cite{jowett06,Salgado:2011wc}. 
The excellent performance was made possible through many improvements in the LHC and the injector chain. 
In particular, the average colliding bunch intensity in 2018 was up to about \unit[$2.3 \cdot 10^8$]{Pb/bunch}, which is more than three times higher than the LHC design value. 
For the next PbPb run in 2021, it is planned to further increase the total LHC intensity through a decrease of the smallest bunch spacing to \unit[50]{ns}, in order to fit 1\,232 bunches in the LHC. 
A further increase of the injected intensity seems difficult without additional hardware in the injector chain~\cite{Coupard:2153863}. 
In the LHC, any increase of ion luminosity is ultimately limited by the risk of quenching magnets, either by secondary beams with the wrong magnetic rigidity created in the collisions~\cite{Klein:2000ba, Jowett:2004rd, Bruce:2007mx, Bruce:2009bg, Schaumann:2015mvg} or by leakage from the halo cleaning by the collimators~\cite{Assmann:2004km, Hermes:2241364, Hermes:2016vhg}.
Mitigation of the secondary beam losses around ATLAS and CMS, using an orbit-bump technique, has been demonstrated
\cite{jowett16-ipac-bfpp} and   additional collimators will be installed in the current long shutdown of the LHC (2019--2020) to allow higher luminosity at IP2~\cite{bahamonde16-ipac-bfpp} and also to raise the total beam intensity limit from collimation losses~\cite{ipac2015:hermesCollTCLD,ApollinariG.:2017ojx}.
Using the predicted beam and machine configuration, the future luminosity performance has been estimated for PbPb and $p$Pb~\cite{Citron:2018lsq}. 
During a one-month run, assuming that the instantaneous luminosity is levelled at the current values around $\unit[6\cdot10^{27}]{cm^{-2}s^{-1}}$, the integrated luminosity per experiment is estimated to be $\unit[3.1]{nb^{-1}}$ for PbPb and $\unit[700]{nb^{-1}}$ for $p$Pb (without levelling), equivalent to NN luminosities of $\Lumi_\text{NN} \approx 0.15$~$\unit{fb^{-1}}$. 

\begin{table}
\centering
\begin{tabular}{rl*{6}{r}}
\toprule
\multicolumn{2}{c}{} & \isotope[16][8]{O} & \isotope[40][18]{Ar} & \isotope[40][20]{Ca} & \isotope[78][36]{Kr} & \isotope[129][54]{Xe} & \isotope[208][82]{Pb} \\ 
\midrule
$\gamma$ & [$10^3$] & 3.76 & 3.39 & 3.76 & 3.47 & 3.15 & 2.96 \\
$\sqrt{s_\text{NN}}$ & [TeV] & 7 & 6.3 & 7 & 6.46 & 5.86 & 5.52 \\
$\sigma_\text{had}$ & [b] & 1.41 & 2.6 & 2.6 & 4.06 & 5.67 & 7.8 \\
$N_b$ & [$10^9$] & 6.24 & 1.85 & 1.58 & 0.653 & 0.356 & 0.19 \\
$\epsilon_n$ & [$\unit{\mu m}$] & 2 & 1.8 & 2 & 1.85 & 1.67 & 1.58 \\
$Z^4$ & [$10^6$] & $4.1\cdot 10^{-3}$ & $0.01$ & $0.16$ & $1.7$ & $8.5$ & $45$\\\midrule
$\widehat \Lumi_\text{AA}$ & [$\unit[10^{30}]{cm^{-2} s^{-1}}$] & 14.6 & 1.29 & 0.938 & 0.161 & 0.0476 & 0.0136 \\
$\widehat \Lumi_\text{NN}$ & [$\unit[10^{33}]{cm^{-2}s^{-1}}$] & 3.75 & 2.06 & 1.5 & 0.979 & 0.793 & 0.588 \\
$\ev{\Lumi_\text{AA}}$ & [$\unit[10^{27}]{cm^{-2}s^{-1}}$] & 8990 & 834 & 617 & 94.6 & 22.3 & 3.8 \\
$\ev{\Lumi_\text{NN}}$ & [$\unit[10^{33}]{cm^{-2}s^{-1}}$] & 2.3 & 1.33 & 0.987 & 0.576 & 0.371 & 0.164 \\
$\int_\text{month} \Lumi_\text{AA} \dd t$ & [$\unit{nb^{-1}}$] & $1.17\cdot 10^4$ & 1080 & 799 & 123 & 28.9 & 4.92 \\
$\int_\text{month} \Lumi_\text{NN} \dd t$ & [$\unit{fb^{-1}}$] & 2.98 & 1.73 & 1.28 & 0.746 & 0.480 & 0.210 \\ 
\bottomrule
\end{tabular}
\caption{LHC beam parameters and performance for collisions from O up to Pb ions, with a moderately optimistic value of the scaling parameter $p=1.5$ introduced in~\cite{jowett_talk_2018,Citron:2018lsq}.  
Here $\sigma_\text{had}$ is the hadronic cross section, $\epsilon_n$ the normalised emittance, and the $Z^4$ factor is provided to indicate the order-of-magnitude enhancement in $\gaga$ cross sections expected in UPCs compared to $pp$ collisions.
Nucleus-nucleus (AA) and nucleon-nucleon (NN) luminosities $\Lumi$ are given at the start of a fill (to simplify the comparison, it is assumed there is no levelling), $\widehat \Lumi$, and as time averages, $\ev{\Lumi}$, with typical assumptions used to project future LHC performance.
Total integrated luminosities in typical one-month LHC runs are given in the last two rows.
}
\label{tab:species1}
\end{table}

In the presently approved CERN planning, it is foreseen to perform another four and a half PbPb runs before the end of LHC Run-4 in 2029, accumulating  $\sim\unit[13]{nb^{-1}}$ in total. 
Furthermore, one short $p$Pb run is planned, as well as one reference $pp$ run. 
No further HI runs have so far been planned after Run-4. These plans would not permit the full exploitation of the BSM possibilities opened up in HI collisions, which require the largest possible integrated luminosities. 
A revised proposal for Runs-3 and 4 and plans to extend the LHC nuclear programme  beyond Run-4 have been formulated~\cite{Citron:2018lsq}.
The additional BSM physics possibilities summarised here complement and reinforce that scientific case.
These studies involve more time spent on $p$Pb runs and also collisions of lighter nuclei, \eg\ Ar, O, or Kr~\cite{jowett_talk_2018,Citron:2018lsq}. Table~\ref{tab:species1} shows estimated beam parameters and luminosity performance for Pb as well as these lighter species. It can be seen that the latter have the potential to reach $\times$(2--15) higher NN luminosities, which would benefit any BSM search based on hard-scattering processes (Section~\ref{sec:hard_scatt}), although the corresponding $\gaga$ luminosities (Section~\ref{sec:UPC}) would be (naively) reduced by a $(Z_\text{PbPb}/Z_\text{AA})^4$ factor. The estimated parameters for a range of lighter ions rely on the assumption that the achievable bunch intensity $N_b$ for a nucleus with charge number $Z$ and mass number $A$ can be scaled from the Pb bunch intensity as $N_b(Z,A) = N_b(82,208) \times (Z/82)^p$, where the power $p = 1.5$ is estimated from previous experience of nuclear beams for the CERN fixed-target experiments and the short run with Xe in the LHC in 2017~\cite{Schaumann:2018qat}. 
It should be noted that these estimates carry a significant uncertainty, since there have been no opportunities to experimentally optimise  these beams for the LHC. 
Furthermore, the integrated luminosity per month in Table~\ref{tab:species1} has been calculated using a simplified model, and no levelling of luminosity, which gives slightly more optimistic values for Pb than the $\unit[3.1]{nb^{-1}}$ stated above, that was simulated with a more detailed and accurate model.
Total integrated luminosities in the range $\unit[0.2\text{--}3]{fb^{-1}}$ are expected depending on the ion-ion colliding system. 
We stress that, if BSM or other physics cases eventually justify it, one can consider running a full ``$pp$ year'' with ions at the LHC, leading to roughly factors of $\times$10 larger integrated luminosities than those listed in Table~\ref{tab:species1}.

%Higher integrated luminosities could be reached in other future machines that are presently on the design stage, such as the High-Energy LHC (HE-LHC)~\cite{Citron:2018lsq} and Future Circular Collider (FCC-hh)~\cite{Benedikt:2018lgy, Benedikt:2018ofy}. 
%It is estimated through simulations that during a 1-month PbPb run, the FCC-hh could produce a total integrated luminosity between $\unit[23]{nb^{-1}}$ and $\unit[65]{nb^{-1}}$, depending on parameters, if leveling is not used and two experiments are taking data. 

\section{Ultraperipheral \texorpdfstring{$\gamma\gamma$}{~~} collisions}
\label{sec:UPC}

In HI collisions, the highly relativistic ions act as a strong source of electromagnetic (EM) radiation, enhanced by the large proton charge number $Z$~\cite{Baltz:2007kq}.
This offers a natural environment in which to observe the photon-initiated production of BSM states with QED couplings.
The cross section for the $\gamma\gamma$ production of any particle $X$ can be calculated within the equivalent photon approximation~\cite{Budnev:1974de} as
\begin{equation}
\sigma_{A_1 A_2 \to A_1 X A_2}
= \int \dd x_1 \dd x_2 \, n(x_1) n(x_2) \widehat \sigma_{\gamma \gamma \to X}
= \int \dd m_{\gamma\gamma}\, \dv{\Lumi_\text{eff}}{m_{\gamma\gamma}} \widehat \sigma_{\gamma \gamma \to X}
\ ,\label{eq:leff}
\end{equation}
where $x_i$ is the longitudinal momentum fraction of the photon emitted by ion $A_i$. 
This factorises the result in terms of a $\widehat\sigma(\gamma\gamma\to X)$ subprocess cross section of a (BSM) system $X$, and fluxes $n(x_i)$ of photons emitted by the ions. 
The latter are precisely determined in terms of the ion EM form factors, and are in particular enhanced by $\propto Z^2$ for each ion, leading to an overall $\sim Z^4$ enhanced production in ion-ion compared to $pp$ collisions (\ie\ a factor of $\sim5\cdot10^7$ for PbPb). 
The experimental signal of UPC processes is very clean with the system $X$ and nothing else produced in the central detector. Moreover, since the virtuality of the emitted photons is restricted to be very small $Q^2 \sim1/R_A^2$, where $R_A$ is the ion radius, the $X$ object is produced almost at rest~\cite{Baltz:2007kq}. The impact parameters $b_\perp$ of UPCs with ions, with $b_\perp\gg 2 R_A$ beyond the range of additional strong interactions, are significantly larger than in the $pp$ case, and the associated gap survival probability is also significantly bigger than for EM proton interactions. 
This latter effect can be accounted for precisely and enters at the $\order{\unit[10]{\%}}$ level in terms of corrections to $\gamma\gamma$ interactions, with rather small uncertainties~\cite{Harland-Lang:2018iur}.
In addition, the background from QCD-initiated production is essentially completely removed by the requirement that the system $X$ and nothing else is seen in the central detector~\cite{Harland-Lang:2018iur}.

\begin{figure}
\includegraphics[height=5.4cm]{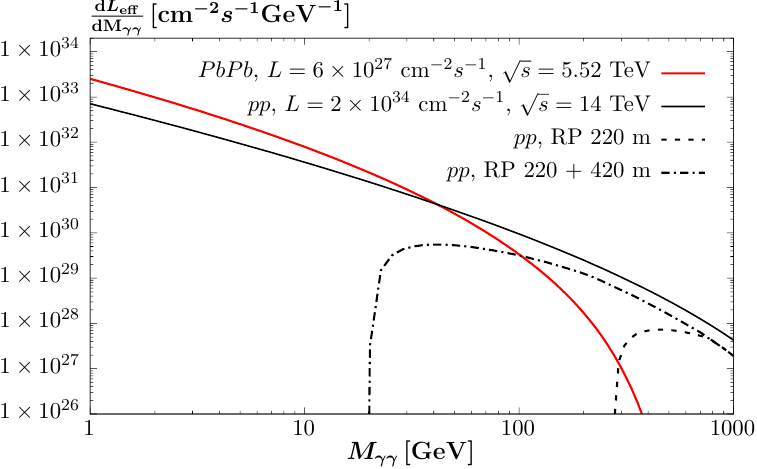}\hfill
\includegraphics[height=5.5cm]{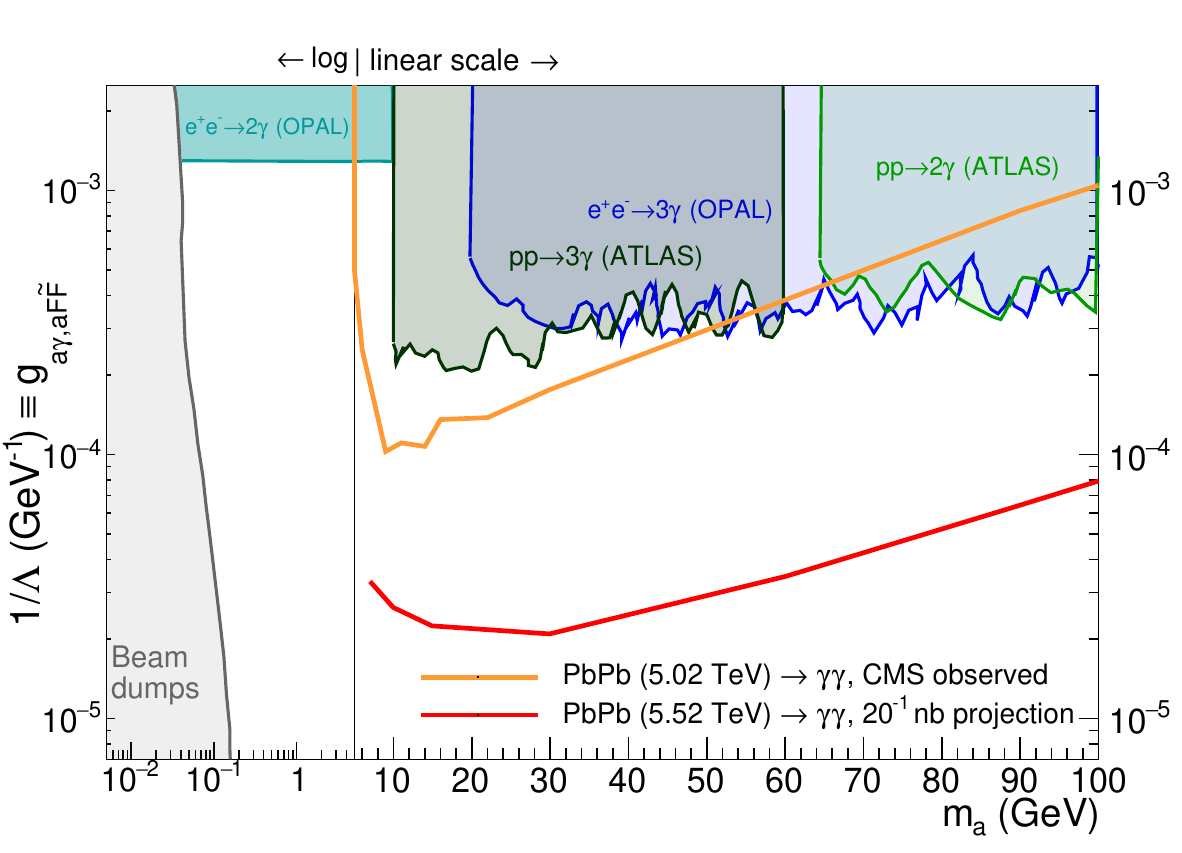}
\caption{Left: Effective $\gamma\gamma$ luminosity vs. photon-fusion mass in ultraperipheral PbPb and $pp$ collisions at the LHC. %, for their nominal $\sqrtsnn$ and instantaneous luminosity values.
In the $pp$ case, the actually ``usable'' $\gaga$ luminosity is also shown with proton tagging at \unit[220]{m} (currently installed) and \unit[420]{m} (proposed). Right: Exclusion limits (95\% confidence level) in the ALP-$\gamma$ coupling ($g_{a\gamma}$) vs. ALP mass ($m_a$) 
plane~\cite{Knapen:2016moh,Sirunyan:2018fhl} currently set in $pp$ and $\epem$ (shaded areas) compared to those from PbPb UPC measurements
(CMS result today~\cite{Knapen:2016moh,Sirunyan:2018fhl}, orange curve; and projections for $\unit[20]{nb^{-1}}$, bottom red curve).}
\label{fig:Leff}
\end{figure} 

A wide program of photon-photon measurements and theoretical work is ongoing in the context of $pp$ collisions at the LHC~\cite{N.Cartiglia:2015gve}, with dedicated proton taggers (Roman Pots, RP) installed inside the LHC tunnel at $\sim$\unit[220]{m} from the ATLAS~\cite{Adamczyk:2015cjy} and CMS~\cite{Albrow:2014lrm} interaction points. In comparison to the $pp$ mode, UPCs with HI offer the distinct advantage of studying such photon-fusion processes in an environment where pileup is absent, forward tagging is unnecessary, and considerably lower masses can be probed. Indeed, 
%the effective $\gamma\gamma$ luminosity in the HI case, due to the $\sim Z^4$ enhancement, is larger than that in $pp$ collisions~\cite{Baltz:2007kq} at lower masses. In addition, 
two-photon processes in $pp$ collisions at high luminosity can only be observed by tagging the forward protons inside the LHC tunnel with geometrical acceptances that bound any central system to have, at least, $m_X \gtrsim \unit[100]{GeV}$. Figure~\ref{fig:Leff} (left) compares the effective $\gaga$ luminosity as a function of $m_{\gaga}$, defined in cross section~\eqref{eq:leff}, for $pp$ and PbPb collisions at their nominal c.m.\ energies and instantaneous luminosities. 
Even after accounting for the reduced beam luminosity in PbPb collisions, the effective  $\gaga$ luminosity is a factor of two higher in PbPb than the (purely theoretical) $pp$ values at low masses. As a matter of fact, taking into account the acceptance in the proton fractional momentum loss $\xi$ of the RP detectors at \unit[220]{m} ($0.02 < \xi <  0.15$)~\cite{Adamczyk:2015cjy,Albrow:2014lrm} and even including proposed RPs at \unit[420]{m} ($0.0015 < \xi < 0.15$)~\cite{Albrow:2008pn}, only PbPb enables studies in the region below $m_X \approx \unit[100]{GeV}$. Running $pp$ at low pileup would cover the low mass region albeit with significantly reduced $\gaga$ luminosities.
%\ifnotarxiv
Various relevant $\gaga$ BSM processes are available in the  {\sc SuperChic}~\cite{Harland-Lang:2018iur} and  {\sc STARlight}~\cite{Klein:2016yzr} Monte Carlo generators, and ion fluxes are also available for any process generated with {\sc MadGraph}~\cite{dEnterria:2009cwl}.
%\fi
%A range of concrete applications of this production mode are outlined below.

\subsection{Axion-like particles}
\label{sec:alp}

Axion-like particles (ALPs) constitute a class of pseudoscalars with couplings to SM fermions or gauge bosons through dimension-5 operators. 
In some cases, they may be Goldstone bosons of an approximate, spontaneously broken, global symmetry.
In this sense they are inspired by original axions arising from the Peccei-Quinn mechanism to solve the absence of CP violation in QCD~\cite{Peccei:1977hh,Wilczek:1977pj}, but in general they do not have to solve the strong CP problem, and are therefore to be understood as purely phenomenological extensions of the Standard Model.
%\footnote{It is worth to remark that, for the purpose of our discussion, a scalar with coupling to photons would behave identically to the pseudoscalar benchmark. The radion in models of warped extra dimensions \cite{Randall:1999ee} is an example.} 
An ALP couples to photons through the operator
%\begin{equation}
$\mathscr{L} \supset \frac{a}{4 f} F_{\mu\nu} \widetilde F^{\mu\nu}\ ,$ %\label{eq:alplag}$
%\end{equation}
where $f$ is the decay constant of the ALP. 
They can be produced through photon-fusion $\gamma\gamma \to a$ or associated $f\bar f \to \gamma a$ production, where the latter tends to be the strongest production mode at electron or proton machines.
In the mass range below about \unit[100]{GeV}, photon fusion in ultraperipheral HI collisions is competitive thanks to the huge $Z^4$ enhancement in the photon luminosity~\cite{Knapen:2016moh} (Fig.~\ref{fig:Leff}, right). 

A second key feature is that the only SM background is light-by-light (LbL) scattering, which is notoriously tiny~\cite{dEnterria:2013zqi}. 
This means that it is crucial that the Lagrangian $\mathscr{L}$ above %~\eqref{eq:alplag} 
provides the dominant coupling of the ALP to the SM: 
Any competing branching ratios to leptons or jets would degrade the reach, as the backgrounds in those final states are unsuppressed.
Evidence and/or observation for LbL scattering in PbPb UPCs has been reported by ATLAS~\cite{Aaboud:2017bwk,Aad:2019ock} and CMS~\cite{Sirunyan:2018fhl}. 
The latter one also provides the best current limits on ALPs in the mass range from $m_a$~=~\unit[5 to 50]{GeV} for coupling to photons only (Fig.~\ref{fig:Leff} right), and  $m_a$~=~\unit[5 to 10]{GeV} for a scenario with hypercharge coupling as well. %~\cite{Sirunyan:2018fhl}.
For a recast of the ATLAS data to a limit on ALPs, see~\cite{Knapen:2017ebd,Citron:2018lsq}.
Given that the higher mass ALPs will be well covered by the regular $pp$ runs, PbPb collisions will likely remain the only choice when searching for ALPs up to $m_a\approx 100$~GeV, though a comparison of the higher-mass reach for lighter ions would be interesting. 
Going below $m_a < \unit[5]{GeV}$ is not possible for ATLAS and CMS, due to trigger and noise limitations in the calorimeters, but the range $m_a \approx \unit[0.5\text{--}5]{GeV}$ can be covered by UPC measurements in ALICE and LHCb, complementing a mass range that Belle~II is also expected to measure reasonably well~\cite{Dolan:2017osp}.
Finally, as more data are gathered, the LbL background will become a limitation.
The limits would therefore benefit substantially if the diphoton invariant mass resolution could be improved, possibly by making use of $\gamma$ conversions.

\subsection{Born-Infeld non-linear QED, non-commutative QED}

The possibility of non-linear Born-Infeld extensions of QED has a long history, first proposed in the 1930s~\cite{Born:1934gh}, they appear naturally in string-theory models~\cite{Fradkin:1985qd}. 
Remarkably, however, the limit on the mass scale of such extensions has until recently been at most at the level of \unit[100]{MeV}~\cite{Ellis:2017edi}. 
The first LHC measurement of LbL scattering in HI collisions~\cite{Aaboud:2017bwk} has enabled to extend the upper limit of non-linear QED modifications by 3 orders of magnitude, up to scales $\Lambda_\text{BI} \gtrsim \unit[100]{GeV}$, which in turn imposes a lower limit of \unit[11]{TeV} on the magnetic monopole mass in the case of a BI extension of the SM in which the $\text{U}(1)_Y$ hypercharge gauge symmetry is realised non-linearly~\cite{Ellis:2017edi}. 
Future LbL measurements in HI UPCs will offer the possibility to further probe Born-Infeld and other non-linear extensions of QED.

Non-commutative (NC) geometries also naturally appear within the context of string/M-theory~\cite{Gomis:2000bn}. 
One consequence of this possibility is that QED takes on a non-Abelian nature due to the introduction of 3- and 4-point functions, leading to observable signatures in the total and differential cross sections of QED processes. 
In~\cite{Hewett:2000zp} it has been demonstrated that non-commutative effects impact $\gaga\to\gaga$ scattering at tree-level, and that a study of its differential cross sections at a photon-collider in the few hundred of GeV range can bound non-commutative scales of order a TeV. 
Somewhat lower limits (in the few hundred GeV range for the NP scale) can be reached through the detailed study of the LbL process accessible in UPCs with ions at the LHC.

\subsection{Other BSM particles}
\label{sec:other-bsm}

%The radion is a BSM scalar particle that naturally arise in models of warped extra dimensions proposed by Randall and Sundrum (RS)~\cite{Randall:1999ee} to explain the large difference between the electroweak and Planck scales. Although interactions of the radion are similar to those of the SM Higgs boson, the existence of a relatively light radion ($m_R\lesssim$~100~GeV) has not been severely constrained so far from the low-mass diphoton Higgs direct searches at LEP and the LHC~\cite{Richard:2017kot}. Being a scalar, axion production benefits highly from the significantly increased photon fluxes in UPC of HIs, and their searches follow closely the ALPs approach...

There are several other possible BSM signals that couple to a pair of photons. It has been argued \eg\ that $\gamma\gamma \to \gamma\gamma$ collisions can be used to search for radions~\cite{Richard:2017kot}, gravitons~\cite{Cheung:1999ja, Davoudiasl:1999di} and unparticles~\cite{Cakir:2007xb}. The UPC signatures would be resonances and/or a non-trivial interference pattern of these new contributions with the SM LbL background. 
The scalar radion would behave identically to the pseudoscalar ALP example discussed in Section~\ref{sec:alp}. Evaluating the search potential requires dedicated studies, in particular to compare with the reach of other studies sensitive to these models, such as the mono-photon searches in standard $pp$ collisions. In the case of unparticles, unitarity and bootstrap bounds must be accounted for as well~\cite{Grinstein:2008qk, Delgado:2009vb, Kathrein:2010ej}. 

Charged supersymmetric (SUSY) particles like sleptons and charginos are also natural targets for $\gamma\gamma$ collisions, especially in the squeezed regime where the standard lepton-plus-missing-$E_T$ searches lose sensitivity. 
Although the parameter space accessible to HI collisions has already been ruled out by LEP
for simplified SUSY scenarios, it may be possible to extract a competitive limit with 
$\gamma\gamma$ collisions from the proton beams~\cite{Beresford:2018pbt, Harland-Lang:2018hmi}.
%and references therein for recent work on this subject. 

Magnetic monopoles necessarily couple strongly to photons~\cite{Dirac:1931kp}. Hence it has been suggested that $\gamma\gamma$ collisions are a natural candidate for monopole searches, either by direct detection~\cite{Kurochkin:2006jr, Epele:2012jn, Reis:2017rvb, Baines:2018ltl}, by the formation of monopolium bound-states~\cite{Epele:2012jn, Reis:2017rvb, Fanchiotti:2017nkk} or via the contribution of virtual monopole loops to LbL scattering~\cite{Ginzburg:1982fk, Ginzburg:1998vb, Abbott:1998mw}. However, these approaches have been criticised for their reliance on perturbative loop expansions in the strong monopole coupling~\cite{Gamberg:1999tv, Milton:2006cp}. Such limitations are circumvented in the production mechanism from classical EM fields discussed next.
%\OG{If desired: A velocity dependent coupling has been proposed as a remedy~\cite{Milton:2006cp, Baines:2018ltl}, based on an extrapolation of small-angle electron-monopole scattering.}

\section{Strong electromagnetic fields}
\label{sec:EM}

\subsection{Magnetic monopoles}
\label{sec:monopoles_schwinger}

\begin{figure}
\hfill\includegraphics[height=5.4cm]{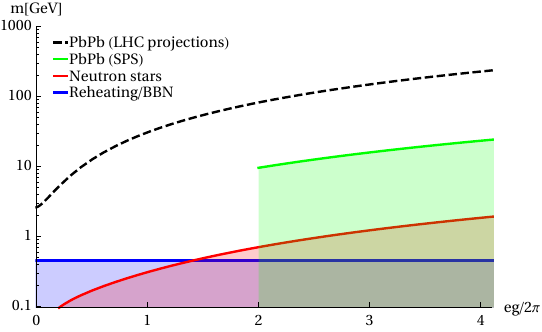}\hfill
\includegraphics[height=5.4cm, width=6.6cm]{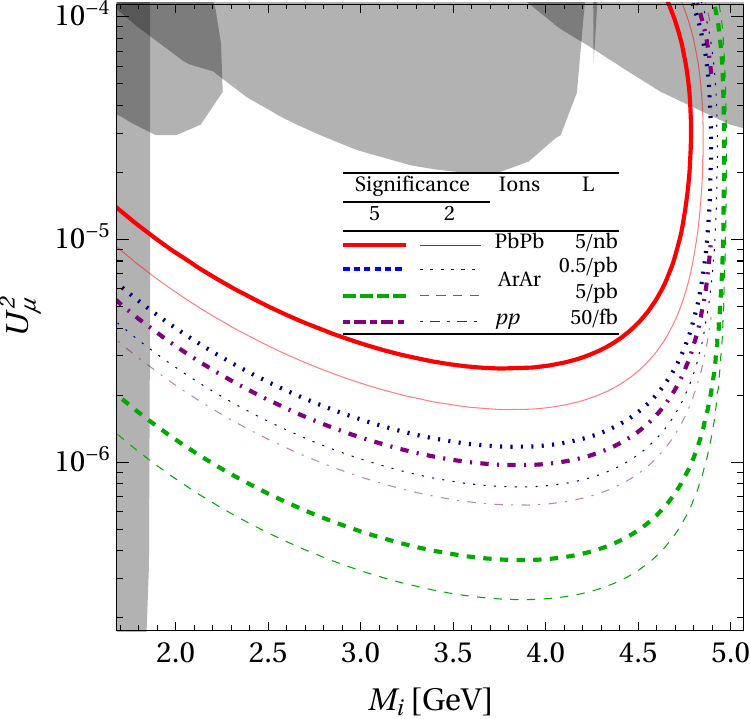}\hfill\strut
\caption{Left: Lower bounds for the magnetic monopole mass ($m$) vs.\ units of magnetic charge ($e\cdot g/2\pi$)~\cite{Gould:2017zwi}. 
Right: Estimated CMS reach for heavy neutrinos, with mass $M_i$ and muon-neutrino mixing angle $U_\mu$, from $B$-meson decays in $pp$, ArAr, and PbPb collisions with equal running time~\cite{Drewes:2018xma}.}
\label{fig:M_LLP_limits}
\end{figure}

There are compelling theoretical reasons for the existence of magnetic monopoles~\cite{Dirac:1931kp, Preskill:1984gd, Polchinski:2003bq}, such as providing a mechanism to explain charge quantisation in the SM.
Consequently, there have been many searches~\cite{Tanabashi:2018oca}, including currently a dedicated LHC experiment, MoEDAL~\cite{Acharya:2017cio}.
Due to the Dirac quantisation condition, magnetic monopoles are necessarily strongly coupled,
%, the magnetic fine structure constant being $\alpha_\text{mag} \approx 34$.
hence perturbative loop expansions for their cross sections cannot be trusted.
In fact, it has been argued that the pair production cross section of semiclassical monopoles~\cite{tHooft:1974kcl, Polyakov:1974ek} in $pp$ or elementary particle collisions suffers from an enormous non-perturbative suppression~\cite{Witten:1979kh, Drukier:1981fq, Papageorgakis:2014dma},
%\begin{equation}
$\sigma_{M\overline M} \propto \mathrm{e}^{-4/\alpha} = 10^{-238}$, 
%\label{eq:suppression}
%\end{equation}
independent of collision energy. 
It is not known if the same suppression applies to point-like elementary monopoles, but if it does~\cite{Goebel:1970nr}, it implies that magnetic monopoles cannot realistically be produced in $pp$ collisions, irrespective of the energy and luminosity of the collider.
The assumptions that led to the exponential cross section suppression %\eqref{eq:suppression} 
do not apply to HI collisions due to the non-perturbatively large magnetic fields that are produced, which are strongest in UPCs~\cite{Huang:2015oca}.
These fields may produce magnetic monopoles by the electromagnetic dual of Schwinger pair production~\cite{Affleck:1981ag}, the calculation of which does not rely on perturbative expansions in the coupling.
To date, there has only been one search for magnetic monopoles in HI collisions, conducted at SPS~\cite{He:1997pj}, which has led to the strongest bounds on their mass~\cite{Gould:2017zwi}.
Searches in HI collisions at the LHC could in principle produce 2--3 orders of magnitude heavier monopoles, directly testing their existence in the hundreds of GeV mass range for the first time (Fig.~\ref{fig:M_LLP_limits}, left).

From the experimental point of view, triggering and tracking constitute challenges for the LHC experiments.
Magnetic monopoles would manifest as highly ionizing particles, and their trajectories in a uniform magnetic field are parabolic.
%\OG{For monopoles produced by strong fields, $v\approx c$ is possible, due to the acceleration from the magnetic field in the collision.}
These are striking features that, on the one hand, help to reject background events to very small levels and, on the other, may cause monopoles to be missed by standard reconstruction algorithms, as a basic assumption of charged-particle tracking is that particle trajectories are helical. 
Given that their production by strong magnetic fields is most likely in UPCs, the usual UPC signature of an almost empty detector would be exploitable to select monopole events.
Alternatively a monopole search can be carried out using passive trapping detectors, exploiting the absolute stability of monopoles as used in the MoEDAL experiment~\cite{Acharya:2017cio}, during the HI running mode.
%This is the method used in the MoEDAL experiment. 
Unlike active detectors, this method gives no direct information about the process that produced the monopole, but it has the advantage that there is no SM background and therefore no risk of a false positive event.

\section{Hard scattering processes}
\label{sec:hard_scatt}

\subsection{Long-lived particles}

%\ifnotarxiv
%\begin{figure}
%\centering
%\includegraphics[width=.4\linewidth]{misidentification}
%\includegraphics[width=.6\linewidth]{B-meson-sensitivity}
%\caption{(left) Example of a signature whose reconstruction can profit from the absence of multiple primary vertices. (right) CMS reach for displaced heavy neutrinos produced with different nucleons with $L_\text{int}=\unit[5.79 \times 10^4,\:7.72\text{ and }10^{-2}]{pb^{-1}}$ for $pp$, ArAr and PbPb, respectively.}
%\label{fig:my_label}
%\end{figure}
%\fi

Many BSM models predict the existence of \emph{long lived particles} (LLPs) that can travel macroscopic distances after being produced, 
cf.\ \eg~\cite{Curtin:2018mvb,Alimena:2019zri}. Their existence is in many cases linked to the solution of fundamental problems in particle 
physics and cosmology, such as the origin of neutrino masses, the DM puzzle, or baryogenesis.
The LLPs usually owe their longevity to a (comparably) light mass, a feeble coupling to ordinary matter, or a combination of both.
If such particles are produced in %head-on 
HI collisions, the feeble interaction allows most of them to leave the quark-gluon plasma unharmed.
Due to the long lifetime, the tracks from their decay into SM particles can easily be distinguished from the large number of tracks that originate from the collision point (a single one, given the absence of pileup when running in the HI mode).
Hence, HI collisions can potentially provide a cleaner environment for LLP searches than $pp$.
The main obstacle is the considerably lower luminosity in HI compared to proton runs, which means that the total number of LLPs produced in the former is always much smaller than in the latter.
However, there are at least three factors that can make the observable number of LLP events competitive~\cite{Drewes:2018xma,Drewes:2019vjy}.

First, due to the absence of pileup, the probability of misidentifying the primary vertex is practically negligible for HI collisions because all tracks originate from a small (fm-sized) region.
This is in contrast to the HL-LHC pileup with proton beams, which leads to a comparable number of tracks as a single PbPb collision~\cite{Apollinari:2015bam, Apollinari:2017cqg} originating from different points in the same bunch crossing and thereby creating a considerable combinatorial track background for displaced signatures. HI collisions entirely remove the problem of identifying the location of the primary vertex, which may be the key to trespass the ``systematics wall'' due to uncertainties in cases where background contamination mostly comes from real (as opposed to misidentified or fake) SM particles.
Although a large track multiplicity is expected to degrade the reconstruction and identification of displaced vertices, the adverse effect of pileup on vertex-finding performance is coming more from the presence of additional primary-interaction vertices than from the sheer number of tracks, as demonstrated by the better $b$-quark tagging performance in $p$Pb compared to $pp$ collisions in $t\bar t$ studies~\cite{Sirunyan:2017xku}.
%using the same algorithm as the standard \p\p\ analysis, and an equal efficiency of correctly tagging $b$-quark-initiated jets, the misidentification rate of light jets is smaller in \p\Pb\ events (\unit[0.1]{\%} vs.\ \unit[0.8]{\%}) in spite of the larger track multiplicity.
%Although those algorithms will have to be retuned to recover a comparable efficiency in the more extreme conditions of high-centrality \Pb\Pb\ collisions, we take it as an indication that at first order pileup affects displaced-particle performance more than track multiplicity.%
%Hence, HI collisions provide a cleaner environment to search for signatures stemming from the decay of LLPs that are heavy enough that the decay products' momenta are not colinear.

Second, absence of pileup allows the detectors to be operated with minimal (zero bias) triggers.
This is an advantage \eg\ in scenarios in which LLPs lead to low-$p_T$ final states.
Third, in addition to the hard scatterings that we focus on here, heavy-ion collisions can offer entirely new production mechanisms that are absent in proton collisions, such as production in the strong electromagnetic fields discussed in Sections~\ref{sec:UPC} and \ref{sec:EM} or
in the thermal processes mentioned in section \ref{sec:thermal}.

In Refs.~\cite{Drewes:2018xma,Drewes:2019vjy}, it has been shown for the specific example of heavy neutrinos
that the zero-bias triggers alone can make searches for typical LLP signatures competitive in HI collisions.
Such heavy-$\nu$ could simultaneously explain the masses of the SM neutrinos and the baryon asymmetry of the universe~\cite{Asaka:2005pn}.
For masses below \unit[5]{GeV}, heavy neutrinos are primarily produced in the decay of $B$-hadrons along with a charged lepton, 
but the lepton $p_T$ is too small to be recorded by conventional $pp$ triggers, making more than \unit[99]{\%} of the events unobservable.
As a result, the observable number of events per running time in PbPb with low-$p_T$ triggers is comparable to that in $pp$ 
collisions with conventional triggers: \unit[5]{${\rm nb}^{-1}$} of lead collisions could improve the current 
limits by more than one order of magnitude in comparison to current bounds.
For a small range of masses over \unit[4]{GeV} the improvement would even be of two orders of magnitude.
If lighter nuclei are used, allowing for higher luminosities (Table~\ref{tab:species1}), then HI collisions can 
yield a larger number of observable LLP events per unit of running time than $pp$ (Fig.~\ref{fig:M_LLP_limits}, right)~\cite{Drewes:2018xma}.
%While it can be argued that ``scouting'' and ``parking'' strategies are expected to recover part of the trigger efficiency for this kind of signatures in $pp$ collisions, the same strategies can also find their use in HI runs, allowing for example to include more decay channels that are out of reach in standard searches.

We should emphasise that we refer to the heavy neutrino example here because it is the only case that has been studied in detail so far. 
It is a very conservative example because it only takes advantage of one of the three factors mentioned above, namely the lower $p_T$ triggers. 
In models that predict an event topology that suffers from backgrounds due to pileup, or LLPs that can be produced through 
one of the new mechanisms mentioned above (such as ALPs) heavy-ion data will have an even bigger impact.

\subsection{Dark photons}

\begin{figure}
\centering
\includegraphics[width=.59\columnwidth]{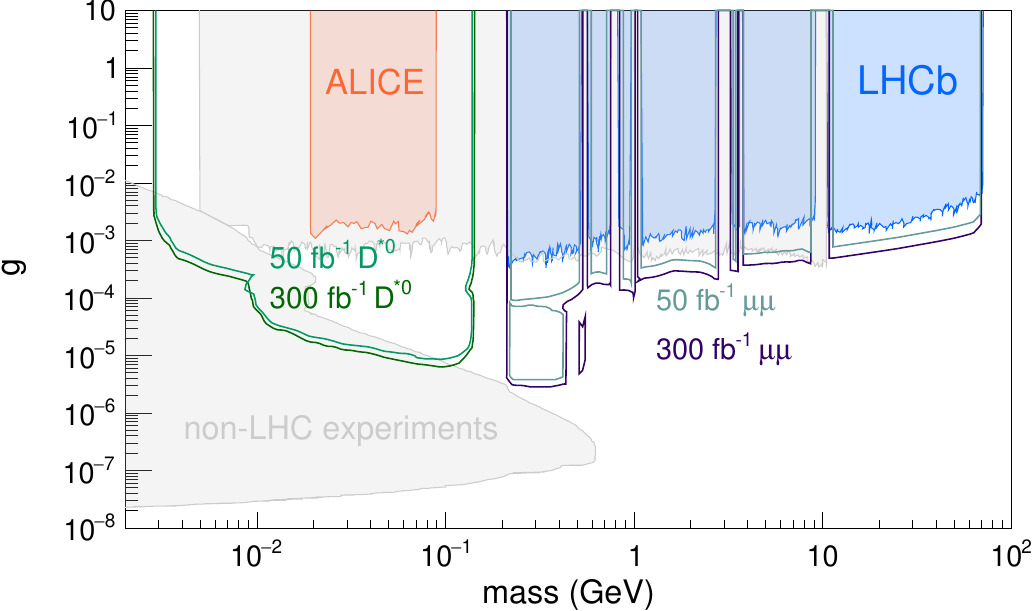}
\includegraphics[width=.40\columnwidth]{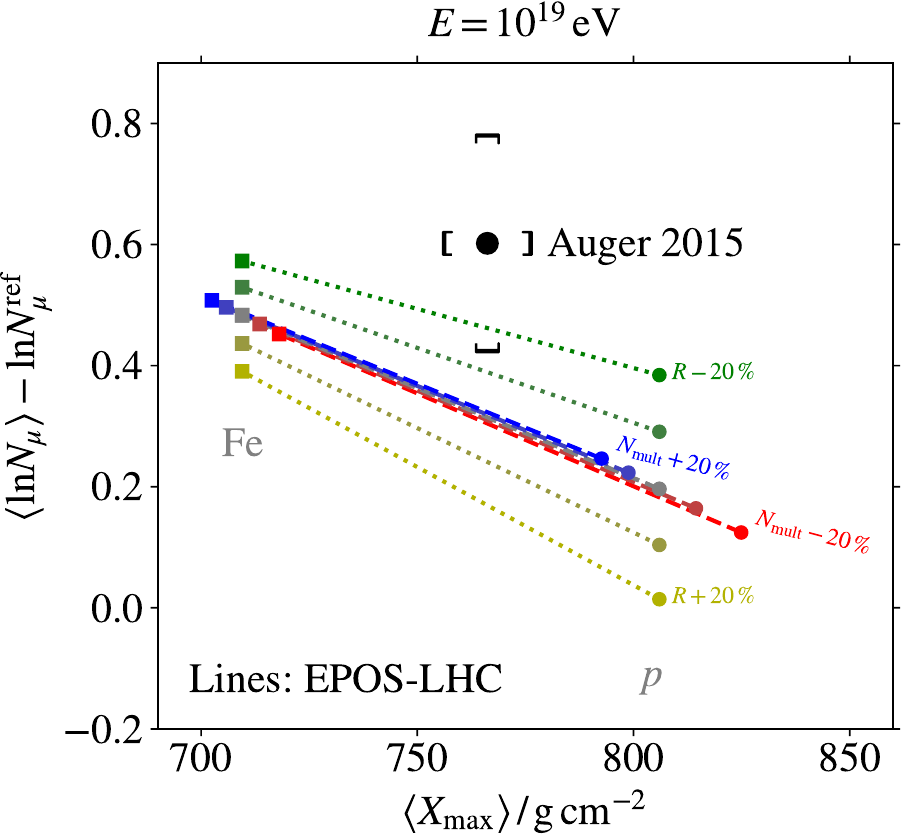}
\caption{Left: 90\% CL exclusion limits of the dark photon mixing parameter $g$ as a function of its mass. 
Red and blue regions show updated projections from measurements at ALICE and LHCb~\cite{Citron:2018lsq}. 
Light grey bands include results from BABAR, KLOE, A1, APEX, NA48/2, E774, E141, E137, KEK, Orsay, BESIII, CHARM,
HPS, NA64, NOMAD, NuCAL, and PS191~\cite{Ilten:2016tkc}. 
% Right: Mass composition of cosmic rays (quantified by the average of the logarithm of their mass number $\log(A)$ 
% as a function of their energy $E$. Model predictions (markers and lines) are compared to data bands from CR muon measurements~\cite{Kampert:2012mx}. 
% Arrows indicate the range covered by $p$O and $pp$ collisions at the LHC, and vertical red lines at the sides 
% indicate the associated experimental uncertainty at low and high CR energies. (Fig. taken from~\cite{Dembinski:2019uta}).
Right: Data-MC deviations in the logarithm of the number of muons produced by a 10$^{19}$~eV CR shower versus its maximum 
depth in the atmosphere ($X_{\rm max}$): Data from Auger~\cite{Aab:2014pza} are compared to MC predictions 
for proton and Fe-ions CR primaries with varying values of the default MC hadron multiplicity $N_{\rm mult}$ 
and the energy fraction $\alpha$ that goes into neutral pions. (Figure taken from~\cite{Baur:2019cpv}).
}
\label{fig:DP_CR}
\end{figure}

The dark photon $A^\prime$ is a hypothetical extra-U(1) gauge boson that acts as a messenger particle between a dark sector, constituted of DM particles, and couples with a residual interaction $g$ to the Standard Model particles.
If the dark photon is the lightest state of the dark sector it can only decay into SM particles.
Typical experimental searches focus on $A^\prime$ decays to dielectrons (if $m_{A'} < 2 m_\mu$), dimuons (for $A^\prime$ masses above twice the muon mass) or dihadrons, and have so far constrained its existence in the mixing parameter $g^2$ versus mass $m_{A'}$ plane.
Collider experiments search for the $A' \to \ell^+\ell^-$ in Dalitz meson decays $\pi^0, \eta, \eta' \to \gamma A'$; meson decays $K \to \pi A'$, $\phi \to \eta A'$, and $D^* \to D^0 A'$;
%Bremsstrahlung process (e ~ Z ~ e ~ ZA 0 with A 0 emitted at very forward direction),
radiative decays of vector-meson resonances %and %initial state radiation
$\Upsilon(3S)$ in BaBar; and $\phi \to e^+e^-$ in KLOE in $e^+e^-$ collisions~\cite{Citron:2018lsq}.
Heavy-ion experiments often feature excellent capabilities for electron and muon identification at low transverse momenta, and for vertexing, leading to competitive searches for low-mass $A'$ from large samples of meson Dalitz decays (see \eg~\cite{Adare:2014mgk} for PHENIX limits in $pp$ and $d$Au collisions at the RHIC collider).
As an example of HI feasibility, ALICE is expected to reach a limit in $g$ of about $10^{-4}$ at \unit[90]{\%} confidence level (CL) for $A'$ masses \unit[20--90]{MeV} with $pp$, $p$Pb and PbPb collisions in Run-3~\cite{Citron:2018lsq} (Fig.~\ref{fig:DP_CR}, left). 
Such limits may eventually be superseded by LHCb and fixed-target experiments~\cite{Ilten:2016tkc}, although any increase in the total HI integrated luminosities, \eg\ running with lighter ions as advocated here, can render the former competitive.

\section{Thermal processes}
\label{sec:thermal}

\subsection{Sexaquarks}
\label{sec:sexaquarks}

The sexaquark $S$ is a hypothesised neutral stable dibaryon $uuddss$ system that can 
%be theoretically formed in the SM and 
account for DM in the universe.
The $S$ would likely have a mass in the range $m_S \approx 2 m_p \pm m_\pi$, and would have escaped detection to date~\cite{Farrar:2017eqq}.
The quark content of the sexaquark is the same as that of the H-dibaryon proposed by R.~Jaffe in 1977~\cite{Jaffe:1976yi}. However, 
the H-dibaryon was assumed to be relatively loosely bound with a weak-interaction lifetime; such a particle has been extensively 
searched for and not found, as discussed in~\cite{Farrar:2017eqq}. Being a flavour singlet, the lightest particle to which 
it could be significantly coupled is the flavour singlet superposition of $\omega$--$\phi$, leading to an estimated size of 
$r_S = \unit[0.1\text{--}0.2]{fm}$~\cite{Farrar:2018hac}. 
If a stable sexaquark exists, it is an attractive DM candidate because the sexaquark-to-baryon density ratio can be predicted by simple statistical arguments in the QGP-hadronisation transition with known QCD parameters (quark masses and $T_\text{QCD}$) to be $\approx 4.5 \pm 1$, in agreement with the observed DM-to-baryon ratio $\Omega_\text{DM}/\Omega_b = 5.3 \pm 0.1$.
This ratio is not modified during the subsequent universe expansion as long as $r_S \lesssim \unit[0.2]{fm}$~\cite{Farrar:2018hac}, thereby evading the counter-arguments against dibaryon DM given in~\cite{Gross:2018ivp, McDermott:2018ofd, Kolb:2018bxv}.

If a stable or weakly-decaying dibaryon exists, its production in HI collisions can be completely predicted as a function of its mass, the temperature $T$, and the local baryon chemical potential $\mu_b$ of the produced QGP.
A rough estimate for the central rapidity region, assuming $m_S = 2 m_p$, $T = \unit[150]{MeV}$, and $\mu_b = 0$, gives $\{\pi:n:S\} \approx \{1:0.01:10^{-4}\}$.
If the entire rapidity range were to come into thermal equilibrium, so that the excess baryon number $B$ of the initial ions is uniformly distributed in rapidity over the final state, in analogy with early universe conditions, it would imply $N_S - N_{\bar S} = \Omega_\text{DM}/\Omega_b (m_p/m_S) (N_B - N_{\bar B})$.
Measuring the dependence of $S$ and $\bar S$ production on $\sqrtsnn$, colliding species, and rapidity would be very revealing and could directly connect DM production in HI collisions to that in the early universe.

Demonstrating that $S$ and $\bar S$ are produced and measuring their production rates is difficult due to the vastly greater abundance of (anti)deuterons with similar mass to $m_S$ and larger scattering and annihilation cross sections in the detector.
Studies are underway to understand the accuracy with which different techniques can identify the production of $S$ and $\bar S$, either exploiting the excellent hadron-identification capability over a wide momentum range in ALICE and LHCb, or the larger acceptance of the multi-purpose ATLAS and CMS detectors. Three basic approaches are being considered~\cite{Farrar:2017eqq}:
\begin{itemize}
\item $S$ particles produced in the primary collision can annihilate with a nucleon in the tracker material and produce a final state with $B=-1$, $S = +2$.
LHC detectors can search for $\bar S p \to K^+ \bar\Lambda^0 $ or $\bar S n \to K^0 \bar\Lambda^0 $.
The $\bar \Lambda^0$ is readily identified; in the absence of an $\bar S$, $\bar\Lambda^0$ production is only consistent with baryon number conservation if the collision is initiated by an antibaryon and the $\bar \Lambda^0$ is accompanied by a baryon.
%The $\bar S n \to K^0 \bar\Lambda^0 $ channel with reconstructed $K^0$ is advantageous here, as that excludes that the neutral is an $n$ or $\Lambda^0$; the decaying $K^0$ is also better for reconstructing the vertex accurately and ensuring that it is an annihilation interaction in the tracker and not some background decay.
Due to the significant penalty for producing a 2-body final state, the rate could be several orders of magnitude greater if the analysis could be extended to events with $>2$ final particles coming from the vertex.
%Excellent hadron-identification over a wide momentum range such as in LHCb and ALICE would be valuable for this purpose.
\item Given the $\order{10^{-2}}$ production rate of $S$ or $\bar S$ relative to single baryons, there may be comparable numbers of events with an $S\bar S$ pair or with just a single $S$ or $\bar S$ produced, with $B$ and strangeness numbers balanced by two (anti)baryons and 0--2 kaons.
%On an event-by-event basis, the particle-ID and hermeticity are not generally sufficient to verify this asymmetry, but 
It may be possible to establish a systematic correlation of missing $\Delta B = \mp 2$; $\Delta S = \pm 2$ on a statistical basis.
\item A population of neutral interacting and/or annihilating particles, distinct from $n$ and $\bar n$ by virtue of having different scattering and annihilation cross sections (and different final states, if that is incorporated into the analysis), is in principle discernible by plotting the rate of such reactions as a function of the tracker material grammage and searching for additional exponential components.
\end{itemize}

%The decay chain $\bar S \to K^+ \bar\Lambda^0 \to K^+ \bar p \pi^+$ can be optimally exploite.d by LHC detectors with excellent hadron-identification capability over a wide momentum range, such as ALICE and LHCb, while the multi-purpose detectors ATLAS and CMS would only be able to rely on invariant masses.

%Although the beam lifetime can be successfully used to set constraints on the density of sexaquark DM captured by Earth, in the event of a positive signal (i.e.\ a significant reduction of beam lifetime with respect to expectations) would be very hard to claim as a hint of new physics, as many known or yet unknown instrumental effects could be invoked as more likely explanations.

\subsection{Magnetic monopoles}

%In Section~\ref{sec:monopoles_schwinger} it was noted that, if magnetic monopoles exist, they may be produced by the strong magnetic fields in ultraperipheral HI collisions. 
For central collisions in which a thermal fireball is created, magnetic monopoles may, in principle, be created thermally. 
Although their microphysical cross sections are not known due to the strong coupling of magnetic monopoles, it seems reasonable to assume that there would exist some production mechanism in a thermal bath containing particles that couple to them (such as photons). 
Thus, if a temperature $T$ is reached in a given HI collision, one would expect to produce monopoles with masses $m \lesssim T$, and an order of magnitude or so heavier when integrated over the luminosity.
Studies based on this production channel would provide an approach to monopole searches independent from, and complementary to, that of production by strong fields (Section~\ref{sec:monopoles_schwinger}). 
However, at LHC energies one would expect production by strong fields to dominate as $T^2\sim \unit[0.3]{GeV^2} \ll g\mathrm{B}\sim \unit[100]{GeV^2}$, where $g=2\pi/e$ is the minimum magnetic charge~\cite{Dirac:1931kp} and B is the magnetic field produced in a typical UPC~\cite{Huang:2015oca}.
The experimental signatures would be as for Section~\ref{sec:monopoles_schwinger}, except that in this case more central HI collisions are favoured.

%\ifnotarxiv

\subsection{Other new physics searches in the QGP}

Studies of other novel QCD phenomena benefit from the larger HI integrated luminosities proposed here:
\begin{itemize}
    \item Various forms of strange-quark matter proposed as DM candidates, such as strange-quark nuggets~\cite{Witten:1984rs} or magnetised quark nuggets~\cite{Abhishek:2018xml}, can form with enhanced rates through thermal production and/or coalescence of partons. The production of any new hypothesised stable multiparton states is therefore expected to be only possible or significantly enhanced out of the hadronizing hot and dense QGP formed in HI collisions.
    \item The absence of CP violation in the QCD sector of the SM is a typical case of theoretical fine-tuning that motivates the existence of new BSM particles, such as the axion~\cite{Peccei:1977hh,Wilczek:1977pj}. An alternative perspective to this problem is provided by finite-temperature studies of the QCD vacuum, whose non-trivial topology leads to the presence of metastable domains with properties determined by the discrete P/CP symmetries. 
Decays of such domains, or classical transitions (sphalerons) among them, in the deconfined QGP phase with restored chiral symmetry can result in local violation of P/CP invariance, leading \eg\ to the so-called ``chiral magnetic effect'' in HI collisions~\cite{Kharzeev:2007jp}. 
\end{itemize}

%\fi

\section{HI input for new physics searches with cosmic rays}
\label{sec:CR}

Beyond colliders, searches for new physics are currently carried out also via cosmic ray (CR) measurements. There are at least two concrete areas where HI data are needed in order to improve the SM theoretical baseline and identify possible BSM signals: (i) precision measurements of antiproton and antinuclei production of relevance for DM searches in space experiments at energies $E_\text{CR} \approx \unit[10^{13}\text{--}10^{15}]{eV}$, and (ii) precision measurements of nuclear effects of relevance for muon production in CR interactions with nuclei in the atmosphere at energies (well) above the LHC range%
\footnote{The LHC $pp$ c.m.\ energy, $\sqrt s = \unit[14]{TeV}$, corresponds to ultrahigh-energy cosmic rays (UHECR) of $E_\text{CR} \approx \unit[10^{17}]{eV}$ colliding with air nuclei at rest in the upper atmosphere.} 
($E_\text{CR} \approx \unit[10^{17}\text{--}10^{20}]{eV}$).

\subsection{Astrophysical DM searches}

Cosmic-ray antiproton and antinuclei have long been considered as potential NP signals, as products \eg\ of DM annihilation, and their detection is a major goal of the AMS-02 experiment on-board the international space station~\cite{Aguilar:2015ooa}.
%The analysis of the antiproton-to-proton ratio~\cite{Aguilar:2016kjl} --- in conjunction with other fluxes or secondary-to-primary ratios --- has revealed excesses at energies between \unit[10 and 20]{GeV} and above \unit[100]{GeV}. 
%While the latter is likely to be explained by reacceleration in supernova remnants~\cite{Cholis:2017qlb}, the former does not point to a simple astrophysical explanation and has been interpreted as a possible signal of DM annihilation in the Galaxy~\cite{Cuoco:2016eej, Cui:2016ppb, Cuoco:2017rxb, Cui:2018klo,Reinert:2017aga}. 
Precise collider measurements of the production cross sections of antiprotons and heavier secondaries in nuclear interactions are crucial ingredients for probing the underlying space propagation~\cite{Genolini:2018ekk,Donato:2017ywo}, and identifying the origin of various excesses observed in the data~\cite{Aguilar:2016kjl} with respect to model predictions~\cite{Cuoco:2016eej, Cui:2016ppb, Cuoco:2017rxb, Cui:2018klo,Reinert:2017aga}. In the absence of new physics, the production of light anti(hyper)nuclei is thought to proceed via thermal hadronisation and 
nucleon coalescence. For instance, a recent measurement of antiproton production in $p$He collisions with the SMOG device of the 
LHCb experiment~\cite{Aaij:2018svt} has significantly improved the antiproton cross-section parametrisation~\cite{Korsmeier:2018gcy} 
used in the interpretation of AMS-02 data~\cite{Cuoco:2019kuu}. 
The cross section for antinucleus production can be parametrised from the ratio of antiproton cross sections in $p$A and $pp$ collisions combined with $A$-dependent coalescence factors $B_A$, that need to be experimentally obtained~\cite{Citron:2018lsq}. 
From the ratios of $B_A$ factors in $p$A and $pp$, one can predict CR flux ratios for a given antinucleus of atomic number $A$. 
The current $B_A$ measurements~\cite{Acharya:2017fvb} are confined to mid-rapidity, and have uncertainties larger than the precision required on the CR flux for DM astrophysical searches. 
Extended LHC running with various ion species is needed to reduce such uncertainties in searches for astrophysical BSM signals via CR antiproton and antinuclei measurements. 

\subsection{Anomalies in ultra-high-energy cosmic-ray showers}

The collisions in the upper atmosphere of the highest-energy CR ever detected, with $E_\text{CR} \approx \unit[10^{20}]{eV}$ corresponding to $\sqrtsnn \approx \unit[400]{TeV}$, are well beyond the reach of foreseeable-future colliders~\cite{dEnterria:2011twh}.
The flux of ultra-high-energy cosmic rays (UHECR) impinging on the earth is very scarce (less than 1 particle per $\unit{km^2}$ per century at the highest energies), and their detection is only possible in dedicated observatories that reconstruct the huge extensive air showers (EAS) of secondary particles that they produce in the atmosphere. 
Measurements of UHECR above LHC energies, $E_\text{CR} \approx \unit[10^{17}\text{--}10^{20}]{eV}$, feature \unit[30--60]{\%} more muons produced at ground and at increasingly larger transverse momenta from the EAS axis, than predicted by all UHECR Monte Carlo (MC) models~\cite{AbuZayyad:1999xa, Aab:2014pza, Aab:2016hkv,Dembinski:2019uta}. Shown in Fig.~\ref{fig:DP_CR} (right) is a representative measurement by the Pierre Auger Observatory~\cite{Aab:2014pza}
showing the data-MC deviation in the number of muons from a 10$^{19}$~eV CR shower versus the maximum depth of the shower in the atmosphere.
The data point is systematically above the EPOS-LHC MC predictions for varying values of relevant model parameters~\cite{Baur:2019cpv}. 
Studies based on PYTHIA~6~\cite{dEnterria:2018kcz} indicate that additional muon production from hard processes, such as from \eg\ jets or heavy-quark 
%(in particular charm~\cite{Engel:2015dxa}) 
decays, do not seem to account for the data-model discrepancy.
The possibility of an additional hard source of muons due to the early production and decay of BSM particles, such as \eg\ electroweak sphalerons~\cite{Brooijmans:2016lfv}, remains an intriguing possibility. 
Solution of the ``muon puzzle'' in UHECR physics requires to reduce the uncertainties on the nuclear effects that remain in the dominant $p$Air (or Fe\,Air) interactions in the top atmosphere.
Dedicated runs of $p$O~\cite{Citron:2018lsq} and light-ion collisions at the LHC are therefore required in order to improve the modelling and tuning of all nuclear effects in the current hadronic MC simulations, before one can consider any BSM interpretation of UHECR anomalies.

%\fi

\section{Summary}

The scientific case for exploiting heavy-ion (HI) collisions at the LHC in searches for physics beyond the Standard Model (BSM) has been summarised. 
A non-comprehensive but representative list of BSM processes accessible with HI at the LHC has been presented based on four underlying mechanisms of production: $\gamma\gamma$ fusion in ultraperipheral collisions, ``Schwinger'' production through strong classical EM fields, hard scattering processes, and thermal production in the quark-gluon-plasma (QGP). 
Such searches provide additional motivations, beyond the traditional QGP physics case, to prolong the HI programme past their currently scheduled end in 2029 (Run-4), in particular running with lighter ion systems, a LHC operation mode that has not been considered so far.
Despite the lower nucleus-nucleus c.m.\ energies and beam luminosities compared to $pp$ collisions, HI are more competitive than the latter in particular in BSM scenarios, whereas in some others they can complement or confirm searches (or discoveries) performed in the $pp$ mode.

Ultraperipheral collisions (UPC) of ions offer, in particular, a unique way to exploit the LHC as an intense $\gamma\gamma$ collider, profiting from the $\sim Z^4$ enhancement factor in their cross sections, providing a clean and well understood environment within which to search for BSM states with QED couplings at masses $m_X \lessapprox \unit[100]{GeV}$ that are otherwise not accessible in the $pp$ mode. 
The UPC discovery potential for new particles, such as axion-like pseudoscalar or radions, and/or new interactions, such as non-linear Born-Infeld or non-commutative QED interactions, is unrivalled in this mass range.
For magnetic monopoles, the huge electromagnetic fields present in HI collisions lead to exponential enhancements of their cross sections and allow for first-principles calculations that are otherwise hindered in similar $pp$ analyses. 
Central HI collisions provide also a propitious environment for searching for a possible stable sexaquark (QCD dark matter candidate). 

In the case of BSM signals produced through hard scatterings, the absence of pileup, %the reduced combinatorial backgrounds,
the improved primary and displaced vertexing, and the lower trigger thresholds of HI compared to $pp$ collisions, provide superior conditions for searches for BSM long-lived particles (LLPs) at low masses: 
An illustrative case has been made based on right-handed neutrinos with $m_\nu \lesssim \unit[5]{GeV}$,
where the higher luminosities attainable with lighter ions lead to a larger number of observable 
LLP events per unit of running time than in $pp$ collisions.
The improved particle identification capabilities and lower $p_T$ thresholds of the ALICE and LHCb experiments 
make them also competitive detectors for dark-photon searches. 
Both LLPs and dark photon searches would benefit from the increased nucleon-nucleon luminosity accessible in collisions with light- and intermediate-ion species. 
Extrapolations based on the current LHC performance indicate that nucleon-nucleon integrated luminosities in the $\unit{fb^{-1}}$ range per month can be easily achieved with lighter ions after the Run-4. 

\subsection*{Acknowledgments}

We thank the participants of the ``Heavy Ions and Hidden Sectors'' workshop, held at Louvain-la-Neuve on 4 and 5 December 2018. This document sprouted from the many lively discussion that we had in that venue, including many people who are not signatories of this document but who had a significant impact in sharpening our ideas about its content. We appreciate in particular Simon Knapen's contribution to Sections~\ref{sec:alp} and~\ref{sec:other-bsm}.
This project has received funding from the European Union's Horizon 2020 research and innovation programme under the Marie Sklodowska-Curie grant agreement 750627.
This work was supported by the F.R.S.-FNRS under the \emph{Excellence of Science} (EOS) project \textnumero\ 30820817 -- be.h.

%\pagebreak

%%\bibliography{New_Physics_and_Heavy_ions-EPPS}
\printbibliography

\end{document}